\documentclass[reprint,pra,aps,floatfix]{revtex4-2}

\usepackage{graphicx}
\usepackage{dcolumn}
\usepackage{bm}
\usepackage{enumerate}
\usepackage{subcaption}

\begin{document}

\title{Understanding the importance of SHAPE to the UK research ecosystem}

\author{H\'el\`ene Draux}
 \affiliation{Digital Science, 6 Briset Street, London EC1M 5NR}
\author{Briony Fane}%
 \affiliation{Digital Science, 6 Briset Street, London EC1M 5NR}%
 \author{Daniel W. Hook}%
 \affiliation{Digital Science, 6 Briset Street, London EC1M 5NR}%
   \author{Philip Lewis}%
 \affiliation{The British Academy, 10-11 Carlton House Terrace, London, SW1Y 5AH}%
  \author{Molly Morgan Jones}%
 \affiliation{The British Academy, 10-11 Carlton House Terrace, London, SW1Y 5AH}%
   \author{Pablo Roblero}%
 \affiliation{The British Academy, 10-11 Carlton House Terrace, London, SW1Y 5AH}%
\author{Juergen Wastl}%
 \affiliation{Digital Science, 6 Briset Street, London EC1M 5NR}%
\author{James R. Wilsdon}%
 \affiliation{Research on Research Institute}
 \affiliation{Department of Science, Technology, Engineering and Public Policy (STEaPP), University College London, 11-20 Capper Street, London, WC1E 6JA}%

\date{\today}

\begin{abstract}
The UK has a long-established reputation for excellence in research across a broad range of fields, but in recent years, there has been greater emphasis on STEM investment and greater recognition of the UK's success in STEM.  This paper examines the relative strengths of SHAPE disciplines and demonstrates that the UK's SHAPE research portfolio outperforms the UK's STEM research, for each international benchmark considered in this work.  It is argued that SHAPE research is becoming increasingly important as a partner to STEM as the widespread use of technology creates societal challenges.  It is also argued that the strength of UK SHAPE is the basis of a strategic advantage for UK research.
\end{abstract}

\maketitle

\section{Introduction}
\label{sec:1} 
The UK has some of the best and most influential arts, humanities, and social science research in the world.  While the UK's STEM (Science, Technology, Engineering and Mathematics) research portfolio is generally internationally regarded, its SHAPE research (Social Sciences, Humanities and the Arts for People and the Economy) does not seem to receive the same level of attention.  Yet, as the current period of apparently exponential technological revolution continues, it is becoming critically important to anticipate the effects of societal interactions with technology by embedding SHAPE at the heart of the development of these new technologies, rather than consigning cultural and societal impacts to afterthought \cite{harari_yuval_2023}.  One such example, perhaps more positively, is the fact that it is clear that SHAPE disciplines played a critical role in the response to and recovery from COVID \cite{shah_covid-19_2021}.  We argue that the interaction between SHAPE disciplines and STEM disciplines can make the research outcomes of each more impactful, maximising the full value of research for societal benefit \cite{wagner_shape_2024}. 

The lack of recognition, both of the role that SHAPE subjects should play, and in many cases already do, to support and enhance STEM disciplines is not uniquely UK-specific and has received significant research attention \cite{benneworth_impact_2016, donovan_introduction_2018}. It is also one that can be readily demonstrated through analysis of, for example, UK government publications.  Successive high-profile UK-government publications appeared between 2019 and 2023, each making the case for investment in the UK's research and innovation system \cite{department_for_science_innovation__technology_international_2019,hm_government_cabinet_office_global_2021,hm_government_cabinet_office_integrated_2023,department_for_science_innovation__technology_uk_2023,department_of_science_innovation__technology_uks_2023,department_for_science_innovation__technology_pioneer_2023}.  Among what amounts to almost 350 pages in these documents, setting out the UK Government's proposed strategic focus for the research and innovation system, the word technology appears 594 times and innovation 383 times; the terms ``science'' and ``scientist'' appear 242 times; and international partnership and collaboration appear 237 times. Ethics, governance, and regulation, words that many would agree are critical to successful delivery of the benefits of technological advances, appear collectively 113 times, and Grand Challenges 57 times. However, the SHAPE disciplines are mentioned just once.

SHAPE as a term is relatively new and was first developed in 2020 by the British Academy, LSE, the Academy of Social Sciences, and Arts Council England \cite{wikipedia_social_2024}.  In the three analyses that we present in the current paper, we see SHAPE through the lens of the Australian and New Zealand Standard Research Classification (ANZSRC) Field of Research (FoR) codes. These codes conform to the Frascati manual \cite{oecd_frascati_2015, australian_bureau_of_statistics_australian_2020} and are used by several governments beyond Australia and New Zealand for coding research \cite{stevenson_data_2023}.  They are also used as the broadest classification scheme in Dimensions \cite{hook_dimensions_2018}, the data source used for the analyses presented here.

This paper presents three analyses that demonstrate the importance of SHAPE subjects in general and, in particular, the influence of the SHAPE disciplines from a UK perspective. Although we present the analysis through the lens of the SHAPE disciplines as compared against the traditional STEM disciplines, this paper is not intended to argue that one set of disciplines is more important than the other. In fact, it is just the opposite.  Undervaluing one set of disciplines over another undermines our ability to make the most of advances in knowledge for societal gain. 

We use bibliometric approaches to make our point and to help start a conversation about how to measure what some might call 'strategic and comparative advantage' in research disciplines. In Section~\ref{sec:2} we give an overview of SHAPE disciplines and STEM disciplines on a national basis using both volume and citation measures.  In Section~\ref{sec:3}, we examine the global influence of UK-based SHAPE research through a network-statistics-based approach, and in Section~\ref{sec:4} we analyse the collaborations between SHAPE subjects and the business community.  We conclude with a brief discussion in Section~\ref{sec:5}.

\section{SHAPE volumes}
\label{sec:2}
\begin{figure*}[!ht]
\includegraphics[width=1.5\columnwidth]{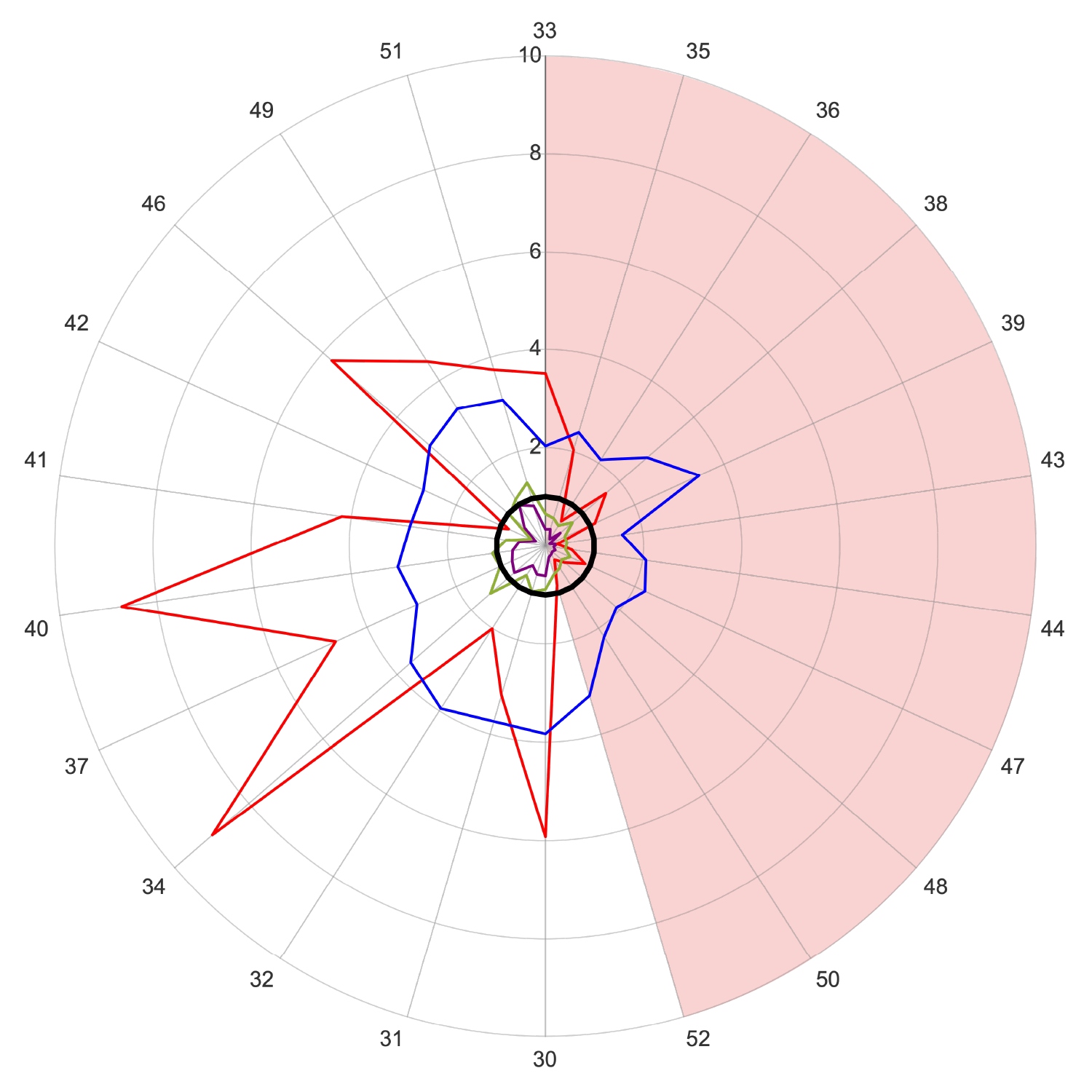}
\caption{Plot of publication output by ANZSRC Field of Research Code in the period 2019-2023 inclusive for major research economies benchmarked to the UK (Black, UK=1); China (Red); US (Blue); Germany (Green); France (Purple).  The SHAPE disciplines have a pink-shaded background.  The STEM disciplines have a white background. \emph{Source: Dimensions from Digital Science.}}
\label{fig:1}
\end{figure*}
Since the end of the 20th Century the research landscape has diversified significantly with a much broader range of countries participating in the global research community. In this section we will seek to understand the relative strengths of STEM versus SHAPE on a national basis.  For this analysis we use the ANZSRC FoR Codes as assigned by Dimensions \cite{porter_recategorising_2023}.  The FoR Codes corresponding to SHAPE subjects are shown in Table~\ref{tab:1}.

\begin{table*}
    \centering
    \begin{tabular}{ccc}
         \textbf{ANZSRC FoR Code}& \textbf{Description}&\textbf{SHAPE/STEM}\\\hline
 30& Agricultural, Veterinary and Food Sciences&STEM\\
 31& Biological Sciences&STEM\\
 32& Biomedical and Clinical Sciences&STEM\\
         33& Built Environment and Design &SHAPE\\
 34& Chemical Sciences&STEM\\
         35& Commerce, Management, Tourism and Services &SHAPE\\
 36& Creative Arts and Writing &SHAPE\\
         37& Earth Sciences&STEM\\
         38& Economics &SHAPE\\
         39& Education &SHAPE\\
         40&  Engineering&STEM\\
         41&  Environmental Sciences&STEM\\
         42&  Health Sciences&STEM\\
         43&  History, Heritage and Archaeology&SHAPE\\
 44& Human Society&SHAPE\\
 45& Indigenous Studies&\textit{Excluded from study}\\
 46& Information and Computing Sciences&STEM\\
 47& Language, Communication and Culture&SHAPE\\
 48& Law and Legal Studies&SHAPE\\
 49& Mathematical Sciences&STEM\\
 50& Philosophy and Religious Studies&SHAPE\\
 51& Physical Sciences&STEM\\
 52& Psychology&SHAPE\\ 
    \end{tabular}
    \caption{ANZSRC FoR Codes defined in 2020. Note that Indigenous Studies is excluded from this study due to a technical limitation on being able to map it into the \textit{Dimensions} dataset - this technical limitatoin is discussed in \cite{porter_recategorising_2023}. }
    \label{tab:1}
\end{table*}

We see directly from Figure~\ref{fig:1} that, relative to the selected developed research economies of the US, China, Germany, and France, the UK performs better over the 5-year period from 2019-2023 in SHAPE disciplines than it does in STEM disciplines by volume. The radar plot in Figure~\ref{fig:1} is normalised such that the UK always scores 1 and the output of other countries is benchmarked relative to the UK. In the pink area, denoting SHAPE subjects, the UK is outperformed only by the US regularly and by China occasionally.  In the STEM areas, the UK is regularly outperformed by the US and China and is regularly challenged by France and Germany. 

To explore this landscape further we examine the trend of a range of metrics.  Each metric represents a specific aspect of a widening sphere of influence.  We begin with the proportion of annual research volume in SHAPE and compare it with the annual research volume in STEM by country---we may think of this metric as a kind of global ``share of voice''.  Research volume is defined to include all of the following types of output: scholarly articles, monographs, edited texts, book chapters, conference proceedings, and preprints.  We then broaden our comparison between SHAPE and STEM at a national level for each of: 
\begin{enumerate}
    \item the proportion of the citations made in a given year to the fractional national attribution of research volume;
    \item the public policy attention in a given year to the fractional national attribution of research volume; and
    \item the patent attention in a given year to the fractional national attribution of research volume.
\end{enumerate}

Each of these trends is designed to give us an insight into a different aspect of SHAPE versus STEM dynamics at an international level.  Research volume tells a simple story of capacity, whereas proportion of global citations to national outputs mixes historic volume with current scholarly attention---establishing an idea of academic relevance.  Public policy attention is a more difficult metric in this context as the policy archive in Dimensions is more Western-centric.  Nevertheless, the coverage provided in Dimensions gives a basis for longitudinal comparison, showing the incremental change in policy attention in a given year to all outputs of a country - again mixing historical volume with current policy attention.  Finally, we examine the annual patent attention trend where, once again, we evaluate the number of patent citations in a given year to the fractional count of papers associated with different countries through their co-authors.  The patent coverage in Dimensions gives good international coverage.  This last metric conflates several traits including each country's propensity to patent (known to be higher in China and the US than in Europe); publication volumes (higher volume provides a higher chance of citation); applicability of research (whether the national focus is on fundamental or highly translatable research); translational capacity in a national context.  Thus, the signals provided by these metrics are not always simple and the message to be drawn from them is not clear cut. 

The temptation with these metrics is to look across countries. Such comparisons may be difficult to justify due to the issues set out above.  However, if we limit our comparison to the difference between the performance of STEM subjects and SHAPE subjects for an individual country, then many of the confounding complexities above are lessened and a consistent picture emerges. 

\begin{figure*}[h]
  \subcaptionbox*{\textit{a}) SHAPE}[.45\linewidth]{%
    \includegraphics[width=\linewidth]{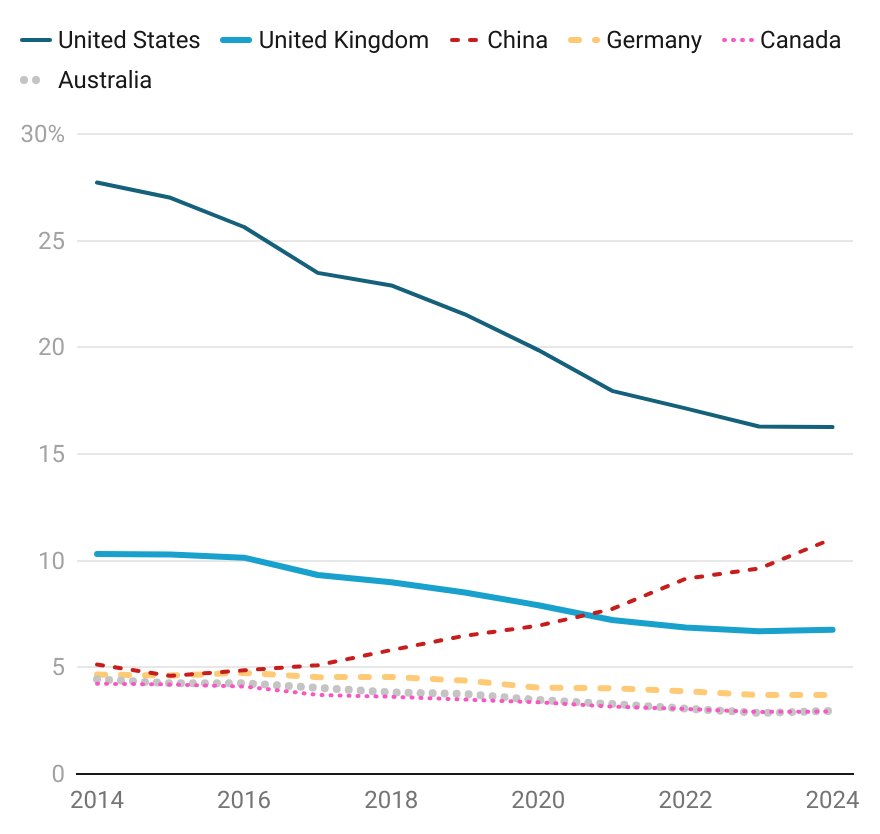}%
  }%
  \hfill
  \subcaptionbox*{\textit{b}) STEM}[.45\linewidth]{%
    \includegraphics[width=\linewidth]{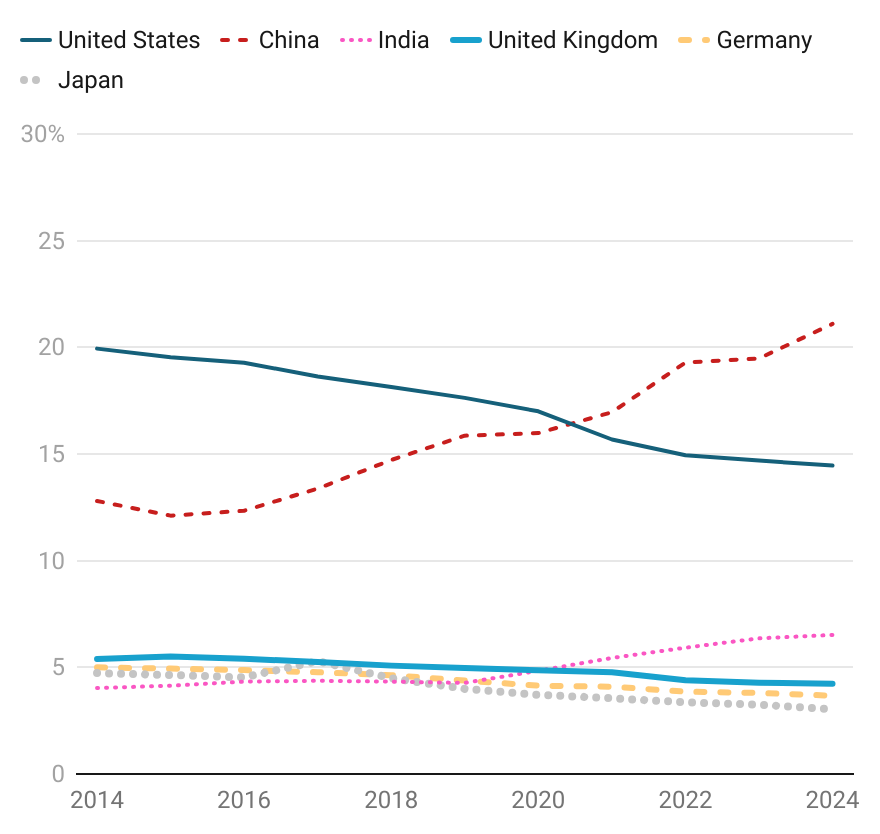}%
  }
  \caption{Proportion of global output using fractional attribution to country in percentage terms from 2014-2024, or ``share of voice'', divided into SHAPE and STEM.  The six most productive research economies for the period 2014-2024 are shown in each case.}
  \label{fig:2}
\end{figure*}

Figure~\ref{fig:2} compares the development of the largest global research economies through each of a SHAPE and STEM lens.  In the STEM picture, China overtook the US as the largest producer of research in around 2021; India is on the rise, and the UK has dropped to fourth position, producing around 4\% of global STEM output.  However, in the SHAPE version of this graph, the US proportion of global output has dropped much more precipitously (almost halving) in a decade at the same time that China has doubled from 5\% of global output to more than 10\%.  Like the US, the UK has also declined in its proportion of global output in the last decade, from just over 10\% to around 7\%, remaining the third largest creator of research content.  It also maintains a comfortable lead ahead of its comparators, Germany, Canada, and Australia. The gap between the UK and its nearest competitors in SHAPE is significantly larger than the gap between it and its nearest competitors in STEM. 

\begin{figure*}[!h]
  \subcaptionbox*{\textit{a}) SHAPE}[.45\linewidth]{%
    \includegraphics[width=\linewidth]{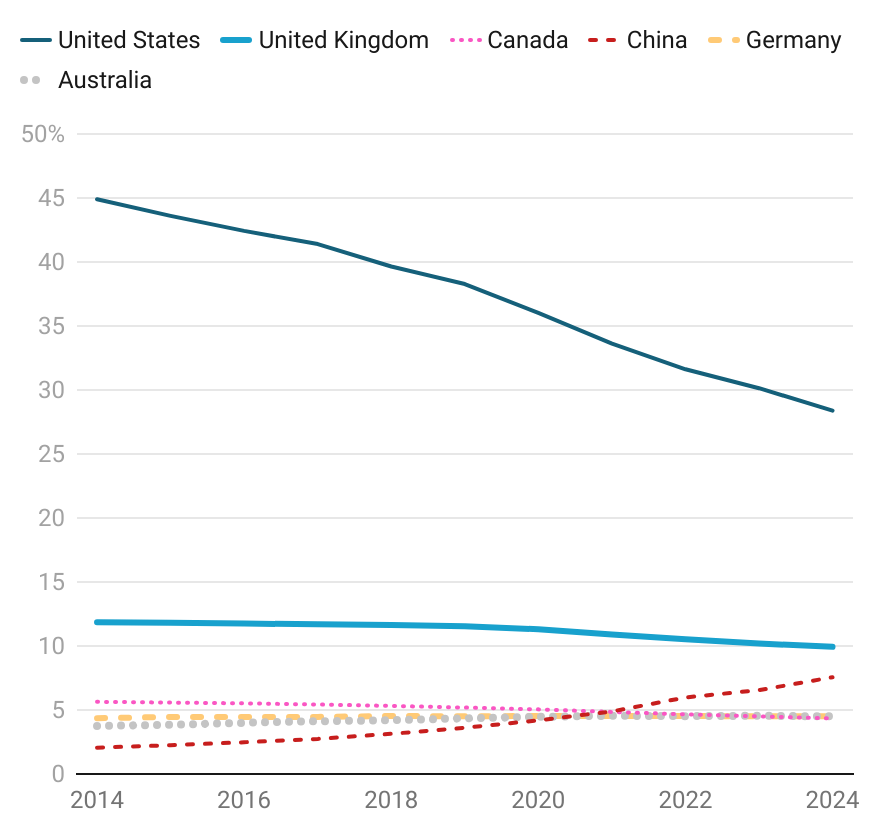}%
  }%
  \hfill
  \subcaptionbox*{\textit{b}) STEM}[.45\linewidth]{%
    \includegraphics[width=\linewidth]{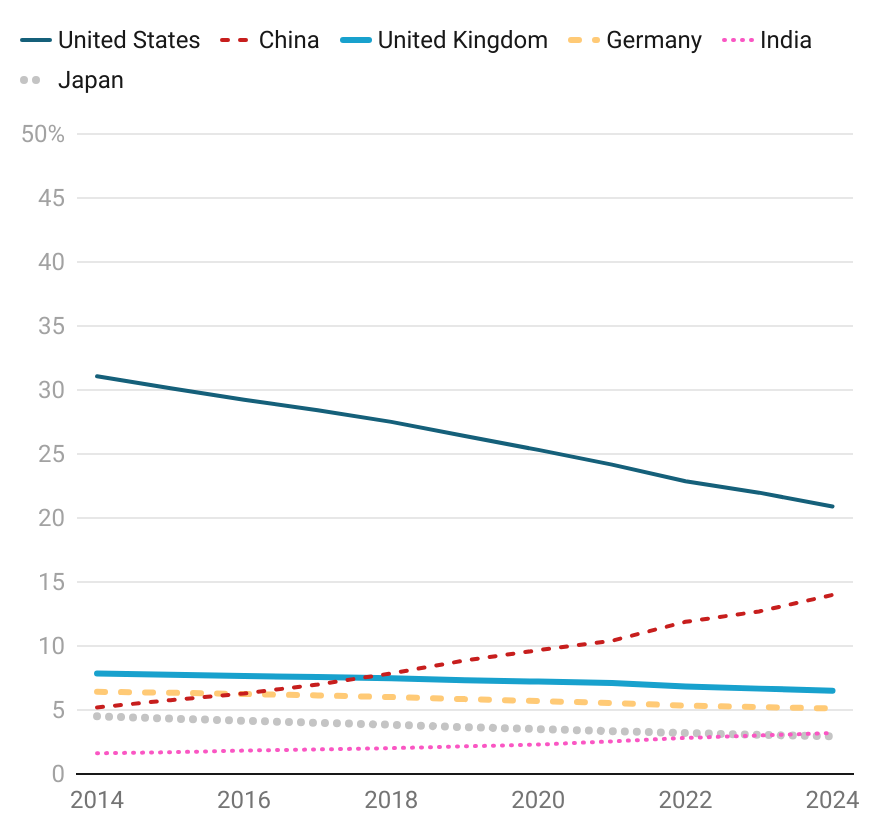}%
  }
  \caption{As Fig.~\ref{fig:2} but with proportion of all citations in a given year to SHAPE/STEM outputs of a given country.}
  \label{fig:3}
\end{figure*}

The next three figures turn from the ``share of voice'' analysis that we have explored in Figure~\ref{fig:2} to ``share of attention'' analyses.  Each analysis explores a different type of attention from scholarly attention (citations), to policy attention (policy documents), industrial or innovation attention (patents).
\begin{figure*}[!h]
  \subcaptionbox*{\textit{a}) SHAPE}[.45\linewidth]{%
    \includegraphics[width=\linewidth]{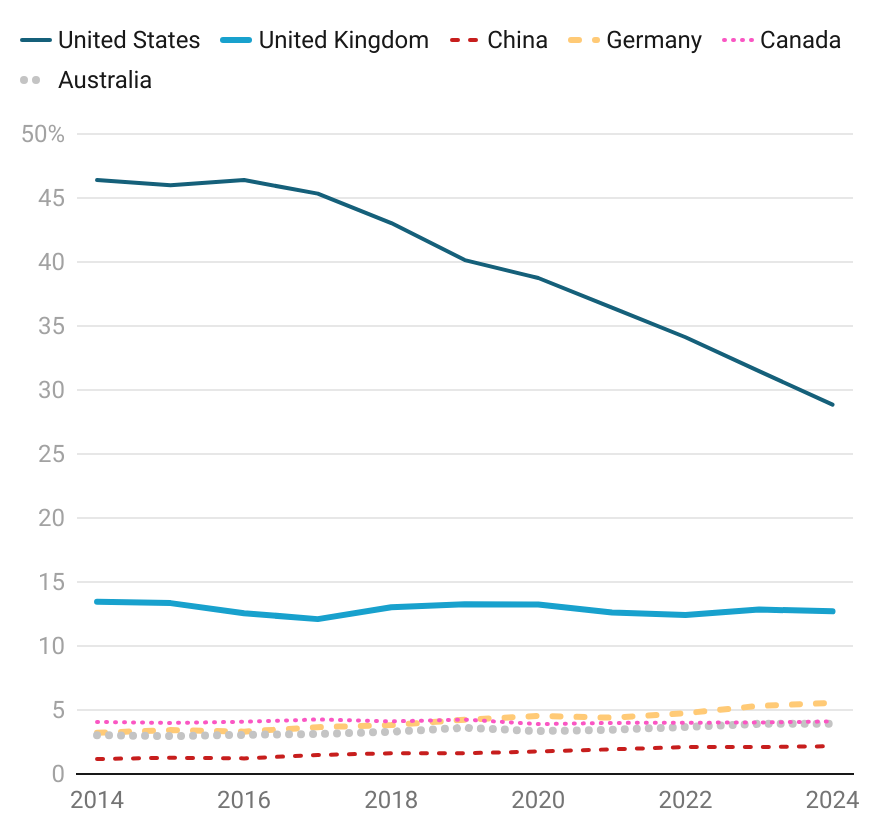}%
  }%
  \hfill
  \subcaptionbox*{\textit{b}) STEM}[.45\linewidth]{%
    \includegraphics[width=\linewidth]{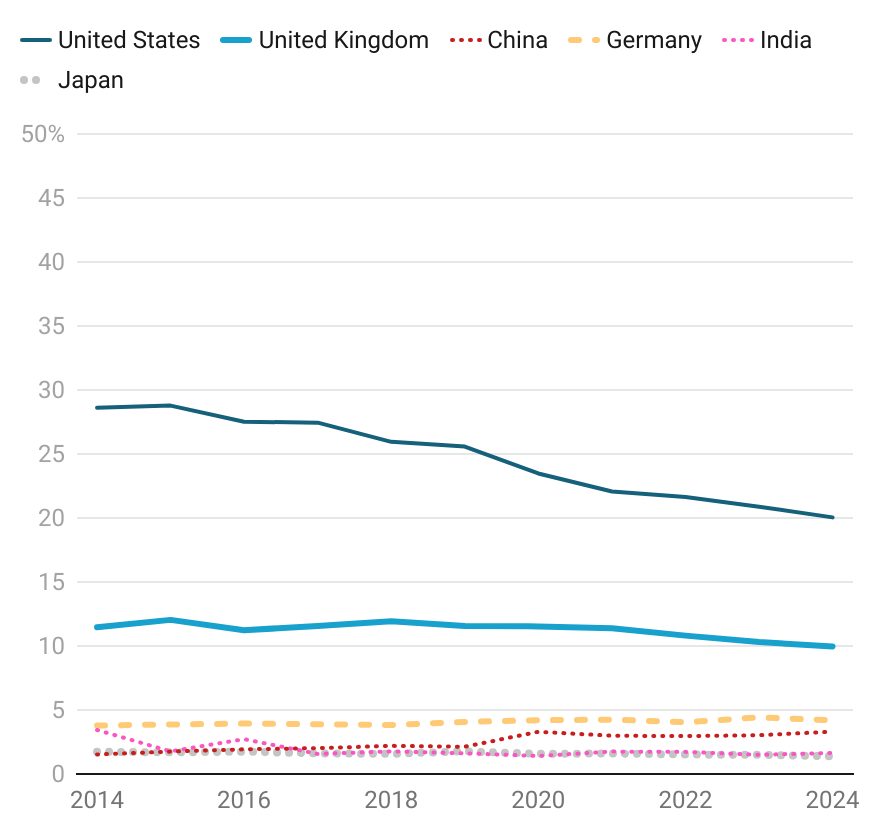}%
  }
  \caption{As Fig.~\ref{fig:2} but with proportion of all policy document citations in a given year to SHAPE/STEM outputs of a given country.}
  \label{fig:4}
\end{figure*}
\begin{figure*}[!h]
  \subcaptionbox*{\textit{a}) SHAPE}[.45\linewidth]{%
    \includegraphics[width=\linewidth]{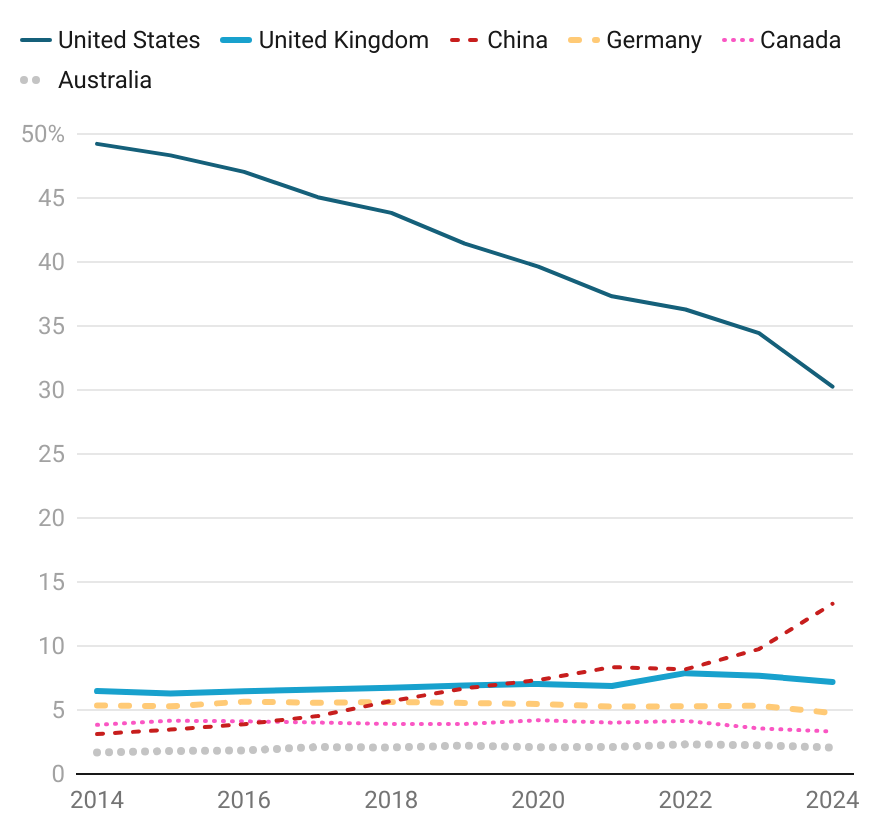}%
  }%
  \hfill
  \subcaptionbox*{\textit{b}) STEM}[.45\linewidth]{%
    \includegraphics[width=\linewidth]{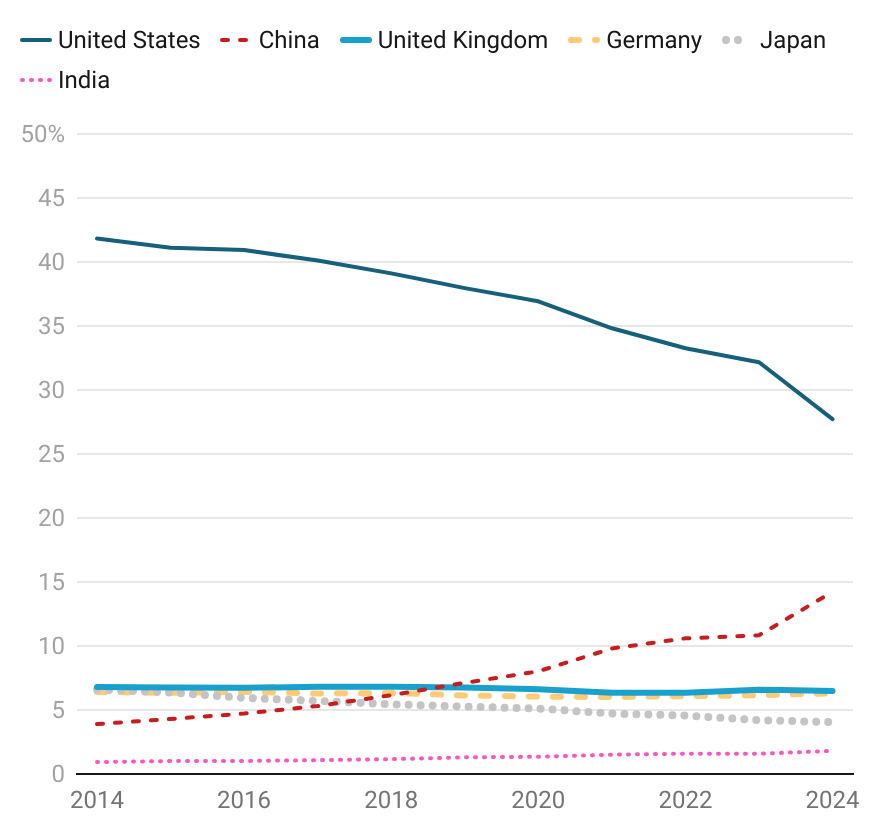}%
  }
  \caption{As Fig.~\ref{fig:2} but with proportion of all patent citations in a given year to SHAPE/STEM outputs of a given country.}
  \label{fig:5}
\end{figure*}

Figure~\ref{fig:3} shows the proportion of scholarly citations made in each year to papers attributed fractionally by country (i.e. if a paper has two co-authors in two different countries, then each country is credited with 50\% of the citations made in that year to the paper).  The countries shown in this plot are kept parallel with those in~\ref{fig:2} for comparison.  In both STEM and SHAPE disciplines, the US maintains a commanding advantage due to its large volume of citable material.  However, while China has made steady progress over the last decade, moving from a 5\% share of global citation attention in 2014, to almost a 15\% share in 2024 in STEM, it has made significantly less progress in SHAPE, despite becoming the world's second largest producer in 2021 (Fig.~\ref{fig:2}).  During the period of this plot, the UK's SHAPE research has always garnered a greater proportion of global scholarly attention than that of its STEM research (c.12\% in 2014 to c.10\% in 2024 in SHAPE compared with c.8\% in 2014 to c.7\% in 2024).  If China is able to maintain its increasing production rate, then it will inevitably lead to a larger proportion of global scholarly attention with time.  However, again, the UK maintains a commanding lead over its other comparators.

As with Figures~\ref{fig:2} and \ref{fig:3} we see a significant decline in the US's share of both global policy document (Fig.~\ref{fig:4}) and patent (Fig.~\ref{fig:4}) attention.  This is again, a symptom of a diversifying world.  However, what is remarkable is that in both cases the UK has managed to maintain a level proportion of each of these types of attention both in SHAPE and STEM.  In Figure~\ref{fig:4} it is less easy to compare between the UK and its non-English-speaking competitors due to intrinsic biases in the dataset, however, it is notable that the UK's SHAPE outputs consistently outperform their STEM counterparts over the decade analysed here. 

Figure~\ref{fig:5} shows a parallel analysis for patent attention to papers--evaluating the proportion of patents each year that cite the fraction of UK-attributed papers: this is what we might call a ``share of patent attention'', which can be thought of as a type of influence measure \cite{jefferson_mapping_2018}.  This graph is specifically unnormalised so that we can engage with the overall share of attention without needing to interpret for the complex landscape that underlies this picture.  Were we to normalise by the volume of publications produced by a country then the resulting metric could be interpreted either as the efficiency of research papers, the level of applied research taking place, or the level of patenting taking place within the country. As the origin of the patent citation is not surfaced in this analysis, and patenting behaviour is extremely different in different countries, a much deeper analysis would be required to understand knowledge flows between countries via patenting, see for example \cite{jaffe_flows_1996}.)  The high proportion of internationally collaborative research is also a confounding factor in analyses of this nature since it is difficult to say that a piece of research is exclusively from a particular country - indeed, we show in the next section that SHAPE research that partners with business (and hence which is more likely to be associated with patent attention) tends to be more highly internationally collaborative.

The performance of the UK in SHAPE and STEM for this metric by the UK is impressive.  For both SHAPE and STEM the UK has maintained its percentage of share of global patent attention even though the US has declined rapidly and China has increased significantly (as part of a greater geographical diversification of the global research economy). Yet, SHAPE research in the UK tends to outperform STEM research in the UK as a proportion of their respective audiences.

These analyses suggest that while the UK's STEM and SHAPE research is declining as a proportion of overall global output, it is remaining of greater scholarly interest to other countries than US counterparts.  It is also remaining more relevant to policy making and innovation (patents), relatively speaking.  However, the UK's SHAPE research consistently outperforms STEM as a proportion of the relevant audiences.  This may make sense as STEM is a highly competitive globally.  Yet, it is clear that China, India and others are not ignoring their investments into SHAPE.  Nonetheless, the UK is maintaining a stronger global position relative to STEM in every metric explored in this section.

\section{SHAPE and influence}
\label{sec:3}
In this section, we examine what has historically been called the \emph{soft power} of the UK as expressed by the international reach of its research connections \cite{nye_soft_2004}.  One mark of the UK's research capability is how preferred it is as a partner in collaborations.  It is appropriate to acknowledge that the UK's imperial history has positioned it to benefit from geopolitical, linguistic, and other infrastructural advantages that go beyond its simply having a long-established research economy. One obvious advantage is that the international language of research continues to be English. While this is a well recognised phenomenon in the STEM disciplines, it may be it introduces even more bias into our analysis here in relation to co-authorship, citation and, as explained below, measures of relative influence. Another advantage this language bias may confer is that researchers may find that studying at UK institutions or collaborating with UK-based colleagues is beneficial. That is unlikely to change rapidly.  The analysis here is not intended to obscure or forget the UK's imperial past, but rather to take it as an unavoidable fact and to understand the nature of the UK's preferred position. Rather than purely examining the advantage of the narrow perspective of preservation, it is also valuable to bear in mind the responsibility that should come with it, and to consider how any remaining advantage could be used positively by the UK for mutual benefit, as a partner and co-creator, and participant in modern cultural diplomacy \cite{nye_public_2008}.  Yet, in a world that is becoming more fragmented, it is questionable whether, even as a positive force, the UK's historic advantages will endure.

A country's soft power in a research context can be thought of as its ability to influence the global research conversation towards its norms and viewpoints. One way of doing this is to produce a large volume of papers.  This is precisely the analysis shown in Figure~\ref{fig:2} corresponding to share of voice.  Another, more subtle way of doing this is to co-author papers with authors from other countries.  In writing the paper, there is a natural exchange of ideas that leads to bi-directional movements of norms and viewpoints.  

To simulate this type of influence, we construct the global network of co-authorship on an annual basis.  But, rather than assigning papers to individual researchers, we assign each paper to the country of the institutions with which co-authors are affiliated.  This leads to a simple weighted network in which there are as many nodes as there are countries that participate in research in a year of interest and connections between those nodes that are weighted by the number of co-authored papers that have been written between collaborators in those countries.  We then calculate a network statistic known as \textit{eigenvector centrality}.  Eigenvector centrality is a well-known network statistic that is often used to infer the relative importance of nodes in a network.  In this case, the importance of a node can be thought of as a proxy for its influence. In the same way that, in a social network, a highly connected individual with deep relationships tends to be highly influential, a country with many papers co-authored with other countries will be highly influential in the global research conversation \cite{hook_pandemic_2023,porterhook2025}.

\begin{figure*}[!h]
  \subcaptionbox*{\textit{a}) SHAPE}[.45\linewidth]{%
    \includegraphics[width=\linewidth]{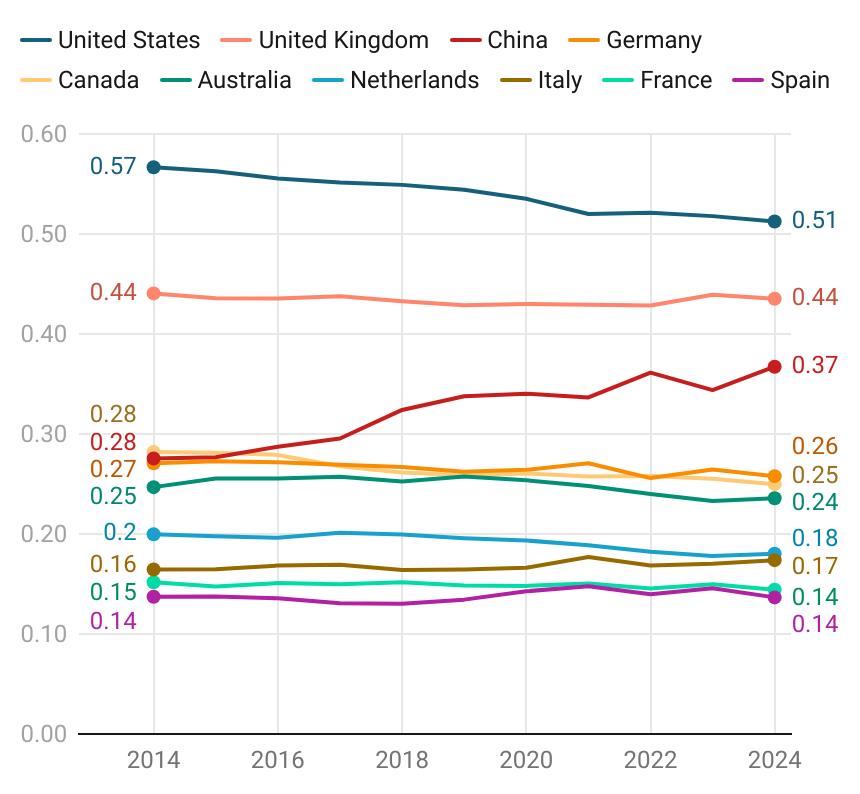}%
  }%
  \hfill
  \subcaptionbox*{\textit{b}) STEM}[.45\linewidth]{%
    \includegraphics[width=\linewidth]{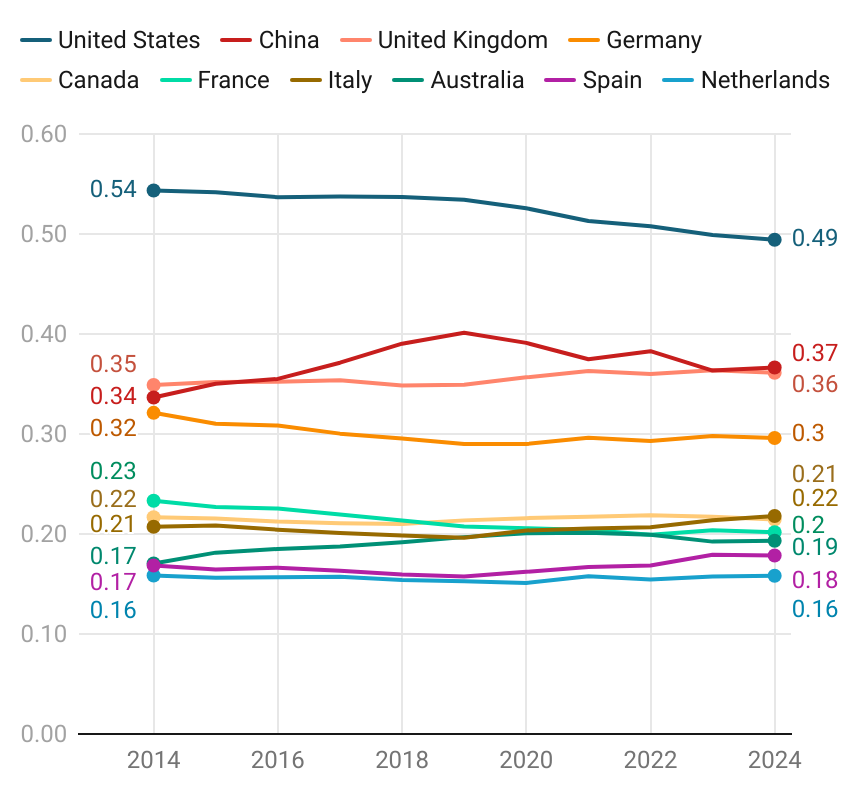}%
  }
  \caption{Eigenvector centrality as an influence measure of countries on the global research narrative.}
  \label{fig:6}
\end{figure*}

On a technical note, we use a Python code to calculate the eigenvector centrality metric - the code uses a version of the algorithm that provides a normalised eigenvector centrality, which has two implications for our analyses - firstly, it is valid to compare the metric across a number of years; secondly, the metric has a ``zero sum'' feel to it, meaning that if one node in the network becomes more influential then other nodes must become relatively less influential.

As in Section~\ref{sec:2} we see that with this metric, the US continues to have the most influential position in global research but that this influence has ebbed with time.  It is noteworthy that while the US appears to be losing out more rapidly in other metrics that we have examined, it continues to be highly influential through the strength of its collaborations as quantified by this metric both in SHAPE and STEM disciplines.  In STEM, the UK has managed to maintain a strong influential position, remaining broadly as influential as China in recent years - a particularly impressive feat given the strength of China's research portfolio in STEM by the other measures that we have reviewed.  However, in SHAPE, the UK is again far more influential relative to its STEM position.  The UK is both significantly more influential than China and much closer to the US's position of influence.  While the UK's position is strong, China's influence in SHAPE is growing rapidly.

As with the metrics in Section~\ref{sec:2} it is clear that even though the UK enjoys a significant and enduring influence in STEM, its influence in the SHAPE world is relatively much more significant.

\section{SHAPE and business collaboration}
\label{sec:4}
In this final section of analysis, we turn our attention to understanding the competitive value of SHAPE disciplines to the UK's industrial base.  For this analysis, we make use of the GRID system in Dimensions, which provides details of research organisations globally, including a classification of type.  This classification includes the following types:
\begin{itemize}
\item Education: An educational institution where research takes place. Can grant degrees and includes faculties, departments and schools.	
\item Healthcare: A health related facility where patients are treated. Includes hospitals, medical centres, health centres, treatment centres. Includes trusts and healthcare systems.	
\item Company: Business entity with the aim of gaining profit.	
\item Archive: Repository of documents, artefacts, or specimens. Includes libraries and museums that are not part of a university.	
\item Nonprofit: Organisation that uses its surplus revenue to achieve its goals. Includes charities and other non-government research funding bodies.	
\item Government: An organisation operated mainly by the government of one or multiple countries.	
\item Facility: A building or facility dedicated to research of a specific area, usually contains specialised equipment. Includes telescopes, observatories, and particle accelerators.	
\item Other: Used in cases where none of the previously mentioned types are suitable.
\end{itemize}

To quantify the level of collaboration with industry, both within a given country and outside that country, we consider publications for which at least one co-author is associated with a Company, and at least one other co-author is affiliated with an institution labelled as Education, as defined above.

\begin{figure}[h]
\includegraphics[width=.95\columnwidth]{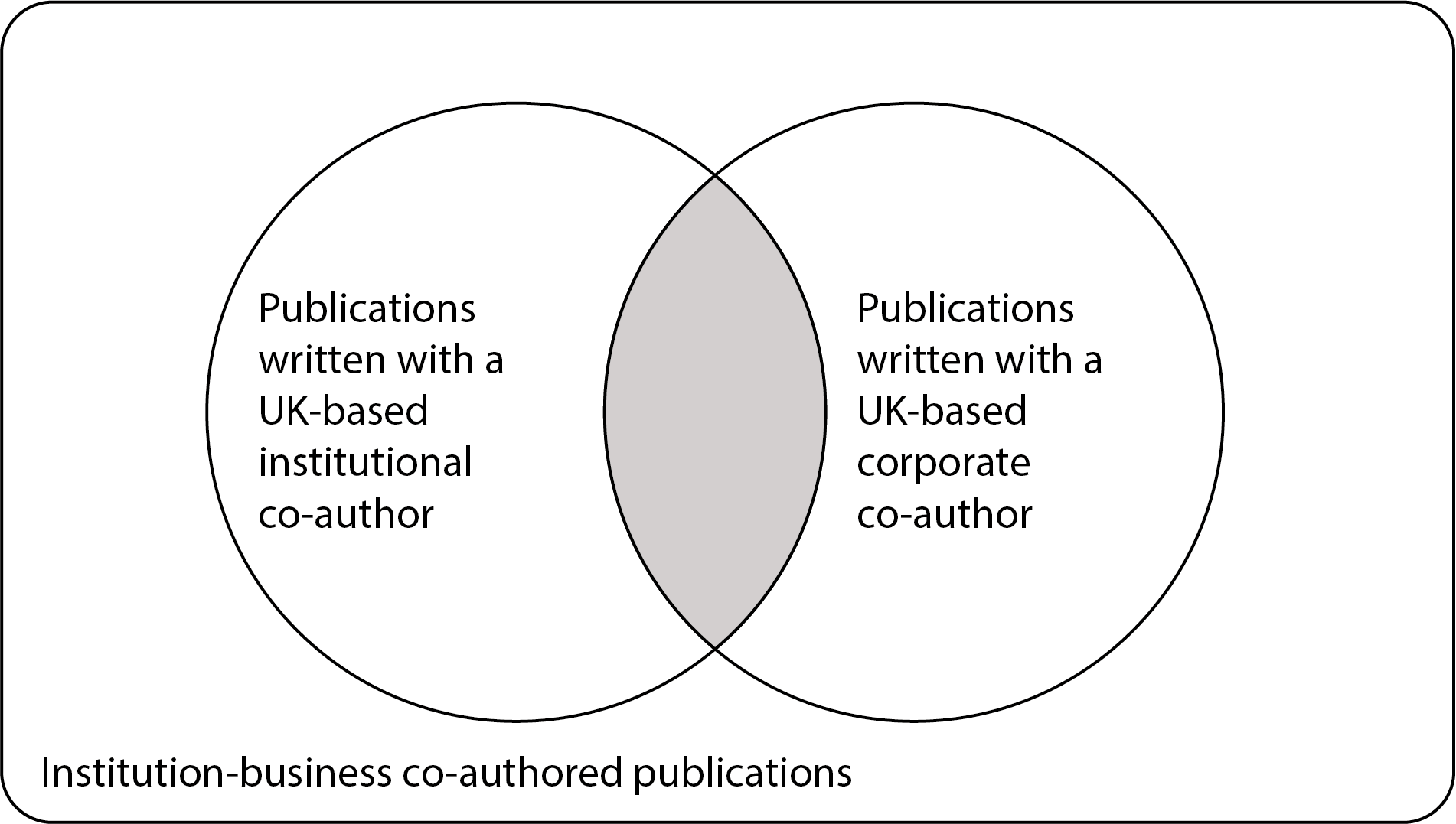}
\caption{Venn diagram to explain the axes of Figures~\ref{fig:8} and \ref{fig:9}.}
\label{fig:7}
\end{figure}

We use Figure~\ref{fig:8} to understand the following figures in this section, in which we have plotted a variety of quantities that map to the different parts of the Venn diagram in Fig,~\ref{fig:7} for a selection of advanced research economies.  The size of each dot in the following figures is determined by the overall volume of institution-business co-authored publications over the period from 2013-2022 respectively.  In each case, the x-axis represents the proportion of the left circle occupied by the grey-shaded region.  For the first case (research volume), this can be interpreted directly as the proportion of a given country’s institutional research that is co-authored with a foreign business (rather than with a business within the country).  Or, for those who think in terms of probability, this percentage corresponds to the probability of finding a paper with an institutionally based author in a chosen country given that the paper has co-authors associated with a foreign company.  Again, in each case, the y-axis represents the analogous ``corporate perspective'' of the same measure.  That is to say that it represents the proportion of the right circle that is occupied by the grey-shaded region, which represents the chance of finding a paper with a co-author at a foreign institution given that there is a co-author at a company in a chosen country of reference. In each plot, countries are coloured according to continent.

We define four notional quadrants that are characterised by specific behaviours:
\begin{enumerate}[i)]
    \item \textbf{International-dominated collaboration} [Top Right]: The quadrant in which, for any chosen paper associated with a given country, there is a high probability of there being a international-corporate co-author collaborating with an in-country institution, or an international-institutional co-author collaborating with an in-country corporate on a given paper. This is a balanced picture but one in which home institutions and home companies tend to be more outward looking.
    \item \textbf{International-institution-driven collaboration} [Top Left] - a region in which international institutions are being sought by companies in a given country - i.e. corporations look abroad for innovation or institutions in their own country are either unaligned or not sufficiently powerful to meet the needs of local companies.
    \item \textbf{International-corporate-driven collaboration} [Bottom Right] - a region in which international companies are being sought by institutions in a given country - i.e. institutions look abroad or are courted from abroad.  This may suggest that institutions cannot find corporate partners to work with locally who are interested in investing in relationships with them.
    \item \textbf{Home-dominated collaboration} [Bottom Left] - The quadrant in which, for any chosen paper associated with a given country, there is a higher probability of there being a home-corporate co-author collaborating with an in-country institution, or a home-institutional co-author collaborating with an in-country corporate on a given paper. This is a balanced picture but one in which home institutions and home companies tend to be more inward looking. This may signal strong innovation culture within a country, but may indicate a level of insularity, a linguistic or geopolitical barrier, or lack of a diversity in the innovation culture of the country.
\end{enumerate}

As with most European countries, the UK finds itself in the outward-looking quarter at the top right of Figure~\ref{fig:8}, which shows the overall (SHAPE plus STEM) international engagement between corporate and institutional co-authorship of publications.  The East Asian countries that have developed their research economies in the last half-century are more introverted.  The US has such a large internal market that it is able to stand alone, but nudges slightly into a mode where its corporate sector is so large, and home to so many multinational companies, that it looks to partner internationally.

\begin{figure*}[!ht]
\includegraphics[width=1.8\columnwidth]{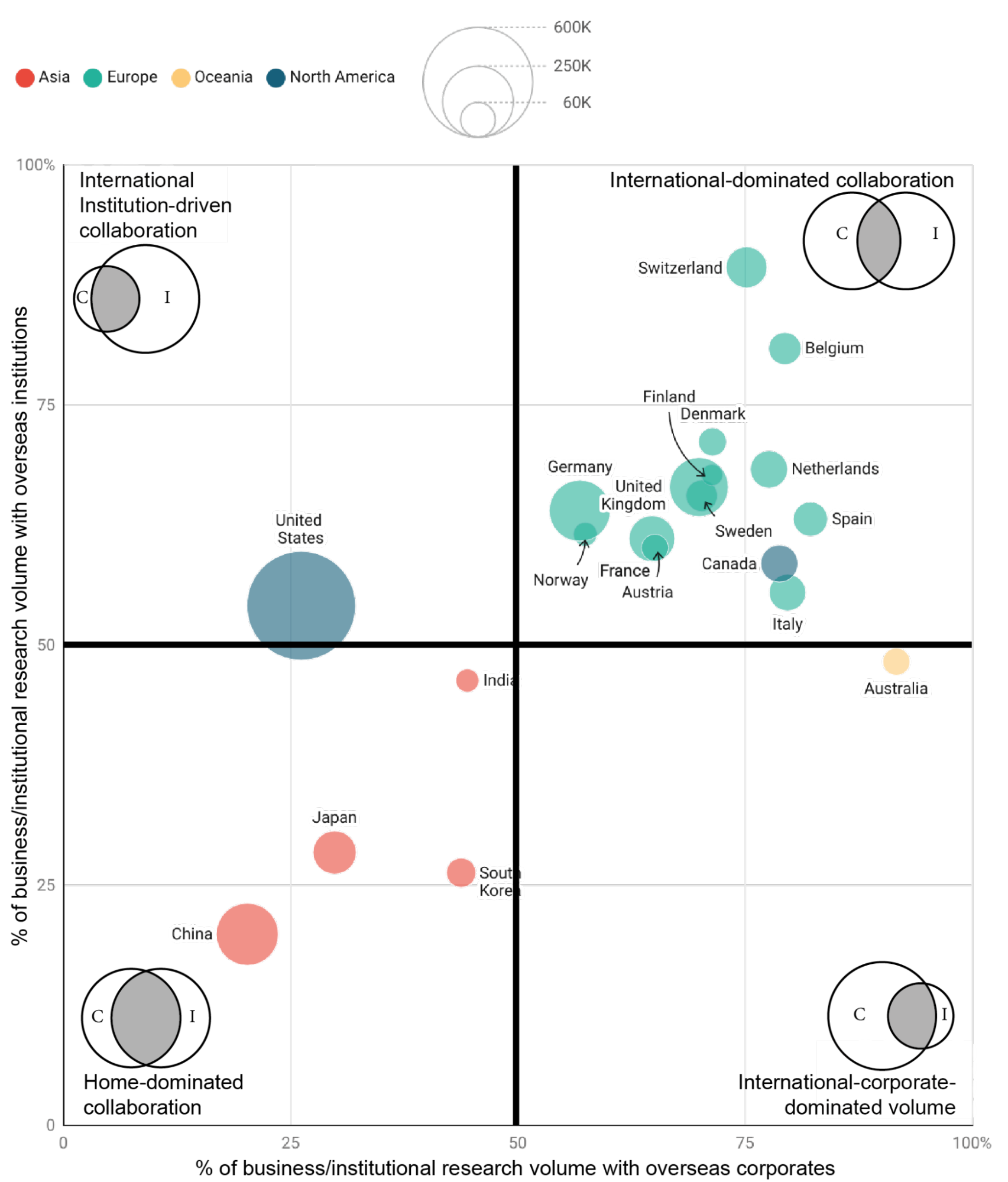}
\caption{Institution and Corporate collaboration volumes for selected countries. Size of point is the total size of the publication volumes.  Colours are continental.  All measures relate to the ten-year period 2013-2022. \emph{Source: Dimensions from Digital Science.}}
\label{fig:8}
\end{figure*}

\begin{figure*}[!ht]
\includegraphics[width=1.8\columnwidth]{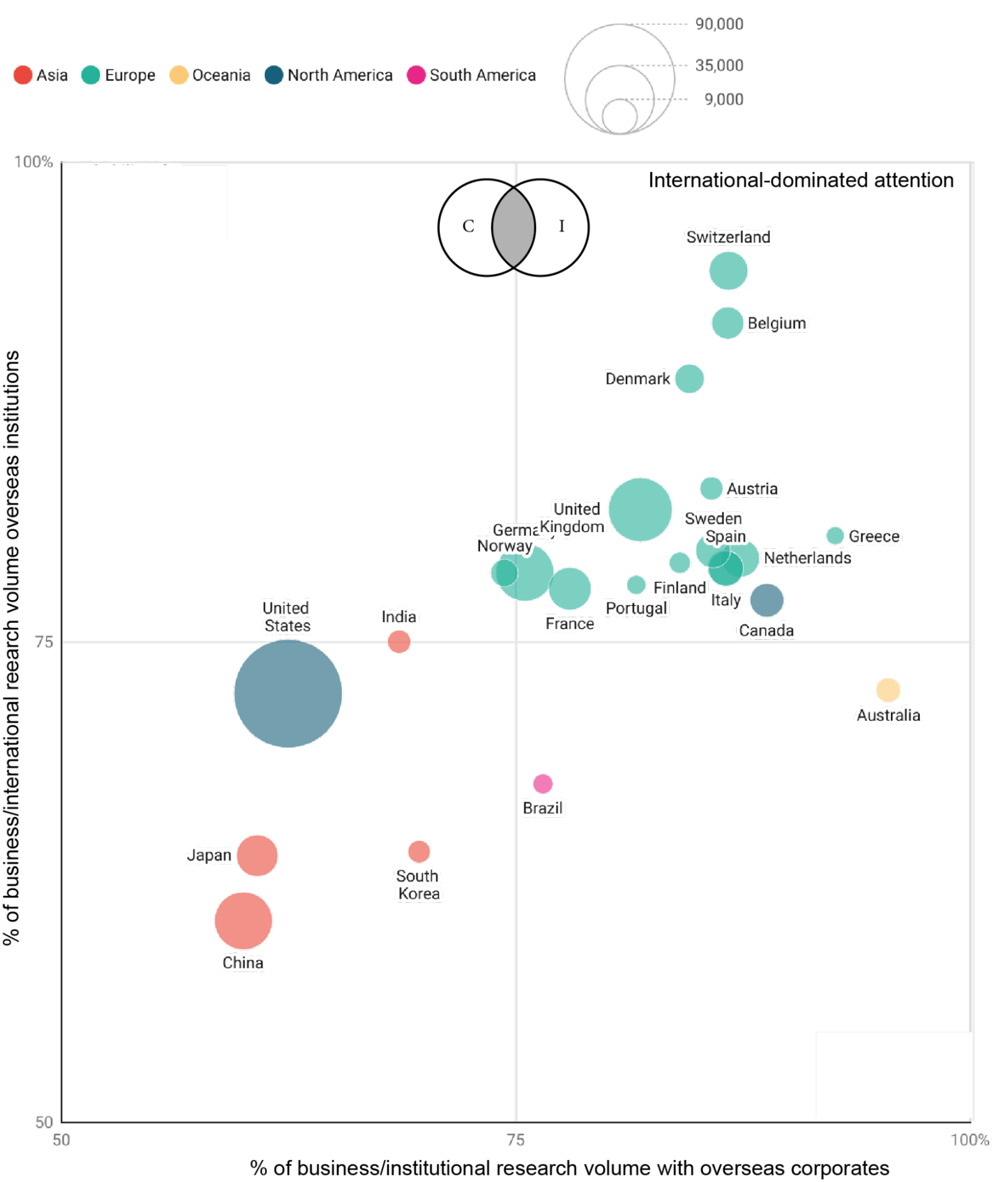}
\caption{As Fig.~\ref{fig:8} but filtered for SHAPE disciplines only.  Note that the scale on the axes of this diagram place it in the upper right quadrant of Fig.~\ref{fig:8} and hence all countries, broadly speaking, are more collaborative internationally in SHAPE disciplines relative to their overall research portfolio. \emph{Source: Dimensions from Digital Science.}}
\label{fig:9}
\end{figure*}

Figure~\ref{fig:9} takes the same analytical frame as Figure \ref{fig:8} but narrows its focus just to SHAPE disciplines.  Firstly, note that the scale of the axes in Figure 18a ranges from 50\%-100\% rather than from 0\%-100\% as in the first figure.  We can conclude straight away that SHAPE research tends to be more international in its range as all the countries shown in Figure \ref{fig:8} migrate to the upper-right quadrant when the subject framing is narrowed to SHAPE.  While many countries retain their approximate relative positioning between Figures \ref{fig:8} and \ref{fig:9}, the UK’s position is improved relative to others.  Note that the size of the UK’s output compared with main comparators such as China, Japan, Germany and France is enhanced, and that both UK business and UK institutions appear to retain their relative overseas attractiveness, even as countries such as France and Germany fall away a little on both dimensions.

Viewed from one perspective, this graph can be interpreted in terms of the UK’s ongoing attractiveness as an international partner for research - both that the UK’s business sector is sought out to participate in research by overseas institutions, but also that the UK’s institutional sector is sought out to participate by overseas business to take part in research collaborations.  The SHAPE disciplines are even more highly prized in this respect as they are proportionally more international in their collaborations both with business and academia.  

\section{Discussion}
\label{sec:5}
We have shown through three different types of scientometric analysis that the SHAPE disciplines in the UK perform consistently as strong, in some cases stronger for their respective audiences, as compared to their STEM counterparts on the measures that we have chosen.  These measures were not specifically chosen in some manner that would give advantage to the SHAPE disciplines.  Rather they were chosen to be generally applicable. Regardless, they paint a picture that the UK is already world-leading in the strength of its SHAPE disciplines. 

In the ``share of voice'' type measure that we suggested, the UK's SHAPE disciplines command a more substantial part of the world's academic conversation than most other countries, with the exception of the US.  This places the UK in a strong position to perform well in other metrics, as we have shown.  However, simply having a strong voice, as measured by output, is not sufficient. We must also show that the research that is being performed in the UK, or with the UK as a significant partner, is of high quality. This means that it receives a high share of scholarly, policy, or translational attention, and is thus recognised and respected internationally.  In these cases, again, the UK performs consistently strongly in SHAPE disciplines than in other disciplinary areas, continuing to hold a more eminent position in the world than many would understand to be the case.

The factors position the UK to have a strong international influence on the research conversation, which we demonstrated using a network-based approach that shows the UK to be more influential in the global SHAPE conversation relative to its ability to influence the international STEM conversation which is much larger in volume.

Finally, we have shown that the UK's research, in general, is highly sought after in a business context both at home and overseas.  Indeed, the UK's SHAPE research is significantly more highly international both when considering academic collaboration and industry collaboration, than the average of UK research. 

These analyses demonstrate that the UK's SHAPE research is under-recognised. Government policy, as evidenced by our opening remarks, has been focused on establishing the UK's status as a ``science superpower'', but to do this will require connecting knowledge across all disciplines. Analyses like these provide the evidence that soft power can and is projected through research. Influence comes from connectedness, and this connectedness is important to bind the entire research system together, regardless of discipline. This is of benefit to the entire community and is in itself a strategic advantage for the UK. The analysis presented here shows the UK already has strong scientific leadership in all disciplines, but particularly so in the SHAPE disciplines. We should not fail to support and maintain that status in future.

\section{Conflict of Interests}
HD, BF, DWH and JW are employees of Digital Science, which owns and operates Dimensions. PL and MMJ are employees of The British Academy, who funded this work.  JRW is an employee of UCL and of RoRI, which is in part funded by Digital Science.

\section{Author Contribution Statement}
Conceptualisation: MMJ, PL; Methodology: JW, JRW, HD, DWH. Formal Analysis: HD, JRW, DWH. Writing-original draft: BF, JW, JRW. Writing-review \& editing: DWH, BF, MMJ, PL. Project Administration: JW. Visualization: DWH, HD.

\section{Data Availability Statement}
All data for this analysis is available at:
https://doi.org/10.6084/m9.figshare.28293137

\section{Acknowledgements}
This work was funded by The British Academy.

\bibliography{BA}

\begin{thebibliography}{23}%
\makeatletter
\providecommand \@ifxundefined [1]{%
 \@ifx{#1\undefined}
}%
\providecommand \@ifnum [1]{%
 \ifnum #1\expandafter \@firstoftwo
 \else \expandafter \@secondoftwo
 \fi
}%
\providecommand \@ifx [1]{%
 \ifx #1\expandafter \@firstoftwo
 \else \expandafter \@secondoftwo
 \fi
}%
\providecommand \natexlab [1]{#1}%
\providecommand \enquote  [1]{``#1''}%
\providecommand \bibnamefont  [1]{#1}%
\providecommand \bibfnamefont [1]{#1}%
\providecommand \citenamefont [1]{#1}%
\providecommand \href@noop [0]{\@secondoftwo}%
\providecommand \href [0]{\begingroup \@sanitize@url \@href}%
\providecommand \@href[1]{\@@startlink{#1}\@@href}%
\providecommand \@@href[1]{\endgroup#1\@@endlink}%
\providecommand \@sanitize@url [0]{\catcode `\\12\catcode `\$12\catcode `\&12\catcode `\#12\catcode `\^12\catcode `\_12\catcode `\%12\relax}%
\providecommand \@@startlink[1]{}%
\providecommand \@@endlink[0]{}%
\providecommand \url  [0]{\begingroup\@sanitize@url \@url }%
\providecommand \@url [1]{\endgroup\@href {#1}{\urlprefix }}%
\providecommand \urlprefix  [0]{URL }%
\providecommand \Eprint [0]{\href }%
\providecommand \doibase [0]{https://doi.org/}%
\providecommand \selectlanguage [0]{\@gobble}%
\providecommand \bibinfo  [0]{\@secondoftwo}%
\providecommand \bibfield  [0]{\@secondoftwo}%
\providecommand \translation [1]{[#1]}%
\providecommand \BibitemOpen [0]{}%
\providecommand \bibitemStop [0]{}%
\providecommand \bibitemNoStop [0]{.\EOS\space}%
\providecommand \EOS [0]{\spacefactor3000\relax}%
\providecommand \BibitemShut  [1]{\csname bibitem#1\endcsname}%
\let\auto@bib@innerbib\@empty
\bibitem [{\citenamefont {Harari}(2023)}]{harari_yuval_2023}%
  \BibitemOpen
  \bibfield  {author} {\bibinfo {author} {\bibfnamefont {Y.}~\bibnamefont {Harari}},\ }\bibfield  {title} {\bibinfo {title} {Yuval {Noah} {Harari} argues that {AI} has hacked the operating system of human civilisation},\ }\href {https://www.economist.com/by-invitation/2023/04/28/yuval-noah-harari-argues-that-ai-has-hacked-the-operating-system-of-human-civilisation} {\bibfield  {journal} {\bibinfo  {journal} {The Economist}\ } (\bibinfo {year} {2023})}\BibitemShut {NoStop}%
\bibitem [{\citenamefont {Shah}(2021)}]{shah_covid-19_2021}%
  \BibitemOpen
  \bibfield  {author} {\bibinfo {author} {\bibfnamefont {H.}~\bibnamefont {Shah}},\ }\bibfield  {title} {\bibinfo {title} {{COVID}-19 recovery: science isn’t enough to save us},\ }\href {https://doi.org/10.1038/d41586-021-00731-7} {\bibfield  {journal} {\bibinfo  {journal} {Nature}\ }\textbf {\bibinfo {volume} {591}},\ \bibinfo {pages} {503} (\bibinfo {year} {2021})},\ \bibinfo {note} {bandiera\_abtest: a Cg\_type: World View Publisher: Nature Publishing Group Subject\_term: SARS-CoV-2, Policy, Society}\BibitemShut {NoStop}%
\bibitem [{\citenamefont {Wagner}\ \emph {et~al.}(2024)\citenamefont {Wagner}, \citenamefont {Rahal}, \citenamefont {Spiers}, \citenamefont {Leasure}, \citenamefont {Verhagen}, \citenamefont {Zhao}, \citenamefont {Li}, \citenamefont {Lu}, \citenamefont {Mills}, \citenamefont {Cooper}, \citenamefont {Degtiareva}, \citenamefont {Epifani}, \citenamefont {McCollum}, \citenamefont {Wyss},\ and\ \citenamefont {{Ref 2021 Lcds Project Team}}}]{wagner_shape_2024}%
  \BibitemOpen
  \bibfield  {author} {\bibinfo {author} {\bibfnamefont {S.}~\bibnamefont {Wagner}}, \bibinfo {author} {\bibfnamefont {C.}~\bibnamefont {Rahal}}, \bibinfo {author} {\bibfnamefont {A.}~\bibnamefont {Spiers}}, \bibinfo {author} {\bibfnamefont {D.~R.}\ \bibnamefont {Leasure}}, \bibinfo {author} {\bibfnamefont {M.}~\bibnamefont {Verhagen}}, \bibinfo {author} {\bibfnamefont {B.}~\bibnamefont {Zhao}}, \bibinfo {author} {\bibfnamefont {L.}~\bibnamefont {Li}}, \bibinfo {author} {\bibfnamefont {Y.}~\bibnamefont {Lu}}, \bibinfo {author} {\bibfnamefont {M.~C.}\ \bibnamefont {Mills}}, \bibinfo {author} {\bibfnamefont {S.}~\bibnamefont {Cooper}}, \bibinfo {author} {\bibfnamefont {E.}~\bibnamefont {Degtiareva}}, \bibinfo {author} {\bibfnamefont {G.}~\bibnamefont {Epifani}}, \bibinfo {author} {\bibfnamefont {B.}~\bibnamefont {McCollum}}, \bibinfo {author} {\bibfnamefont {R.}~\bibnamefont {Wyss}},\ and\ \bibinfo {author} {\bibnamefont {{Ref 2021 Lcds Project Team}}},\ }\href {https://doi.org/10.5871/shape/9780856726866.001}
  {\emph {\bibinfo {title} {The {SHAPE} of {Research} {Impact}}}},\ \bibinfo {type} {Tech. Rep.}\ (\bibinfo  {institution} {The British Academy},\ \bibinfo {year} {2024})\BibitemShut {NoStop}%
\bibitem [{\citenamefont {Benneworth}\ \emph {et~al.}(2016)\citenamefont {Benneworth}, \citenamefont {Gulbrandsen},\ and\ \citenamefont {Hazelkorn}}]{benneworth_impact_2016}%
  \BibitemOpen
  \bibfield  {author} {\bibinfo {author} {\bibfnamefont {P.}~\bibnamefont {Benneworth}}, \bibinfo {author} {\bibfnamefont {M.}~\bibnamefont {Gulbrandsen}},\ and\ \bibinfo {author} {\bibfnamefont {E.}~\bibnamefont {Hazelkorn}},\ }\href {https://doi.org/10.1057/978-1-137-40899-0} {\emph {\bibinfo {title} {The {Impact} and {Future} of {Arts} and {Humanities} {Research}}}}\ (\bibinfo  {publisher} {Palgrave Macmillan UK},\ \bibinfo {address} {London},\ \bibinfo {year} {2016})\BibitemShut {NoStop}%
\bibitem [{\citenamefont {Donovan}\ and\ \citenamefont {Gulbrandsen}(2018)}]{donovan_introduction_2018}%
  \BibitemOpen
  \bibfield  {author} {\bibinfo {author} {\bibfnamefont {C.}~\bibnamefont {Donovan}}\ and\ \bibinfo {author} {\bibfnamefont {M.}~\bibnamefont {Gulbrandsen}},\ }\bibfield  {title} {\bibinfo {title} {Introduction: {Measuring} the impact of arts and humanities research in {Europe}},\ }\href {https://doi.org/10.1093/reseval/rvy019} {\bibfield  {journal} {\bibinfo  {journal} {Research Evaluation}\ }\textbf {\bibinfo {volume} {27}},\ \bibinfo {pages} {285} (\bibinfo {year} {2018})}\BibitemShut {NoStop}%
\bibitem [{\citenamefont {{Department for Science, Innovation \& Technology and Department for Business, Energy \& Industrial Strategy}}(2019)}]{department_for_science_innovation__technology_international_2019}%
  \BibitemOpen
  \bibfield  {author} {\bibinfo {author} {\bibnamefont {{Department for Science, Innovation \& Technology and Department for Business, Energy \& Industrial Strategy}}},\ }\href {https://www.gov.uk/government/publications/uk-international-research-and-innovation-strategy} {\bibinfo {title} {International {Research} and {Innovation} {Strategy}}} (\bibinfo {year} {2019})\BibitemShut {NoStop}%
\bibitem [{\citenamefont {{HM Government Cabinet Office}}(2021)}]{hm_government_cabinet_office_global_2021}%
  \BibitemOpen
  \bibfield  {author} {\bibinfo {author} {\bibnamefont {{HM Government Cabinet Office}}},\ }\href@noop {} {\bibinfo {title} {Global {Britain} in a competitive age: {The} {Integrated} {Review} of {Security}, {Defence}, {Development} and {Foreign} {Policy}}} (\bibinfo {year} {2021})\BibitemShut {NoStop}%
\bibitem [{\citenamefont {{HM Government Cabinet Office}}(2023)}]{hm_government_cabinet_office_integrated_2023}%
  \BibitemOpen
  \bibfield  {author} {\bibinfo {author} {\bibnamefont {{HM Government Cabinet Office}}},\ }\href {https://www.gov.uk/government/publications/integrated-review-refresh-2023-responding-to-a-more-contested-and-volatile-world} {\bibinfo {title} {Integrated {Review} {Refresh} 2023: {Responding} to a more contested and volatile world}} (\bibinfo {year} {2023})\BibitemShut {NoStop}%
\bibitem [{\citenamefont {{Department for Science, Innovation \& Technology}}(2023{\natexlab{a}})}]{department_for_science_innovation__technology_uk_2023}%
  \BibitemOpen
  \bibfield  {author} {\bibinfo {author} {\bibnamefont {{Department for Science, Innovation \& Technology}}},\ }\href@noop {} {\bibinfo {title} {The {UK} {Science} and {Technology} {Framework}: taking a systems approach to {UK} science and technology}} (\bibinfo {year} {2023}{\natexlab{a}})\BibitemShut {NoStop}%
\bibitem [{\citenamefont {{Department of Science, Innovation \& Technology and Foreign, Commonwealth \& Development Office}}(2023)}]{department_of_science_innovation__technology_uks_2023}%
  \BibitemOpen
  \bibfield  {author} {\bibinfo {author} {\bibnamefont {{Department of Science, Innovation \& Technology and Foreign, Commonwealth \& Development Office}}},\ }\bibfield  {title} {\bibinfo {title} {The {UK}’{S} {International} {Technology} {Strategy}},\ }\href@noop {} {\  (\bibinfo {year} {2023})}\BibitemShut {NoStop}%
\bibitem [{\citenamefont {{Department for Science, Innovation \& Technology}}(2023{\natexlab{b}})}]{department_for_science_innovation__technology_pioneer_2023}%
  \BibitemOpen
  \bibfield  {author} {\bibinfo {author} {\bibnamefont {{Department for Science, Innovation \& Technology}}},\ }\href@noop {} {\emph {\bibinfo {title} {Pioneer: global science for global good}}},\ \bibinfo {type} {Tech. Rep.}\ (\bibinfo {year} {2023})\BibitemShut {NoStop}%
\bibitem [{\citenamefont {Wikipedia}(2024)}]{wikipedia_social_2024}%
  \BibitemOpen
  \bibfield  {author} {\bibinfo {author} {\bibnamefont {Wikipedia}},\ }\href {https://en.wikipedia.org/w/index.php?title=Social_sciences,_humanities_and_the_arts_for_people_and_the_economy&oldid=1251138241} {\bibinfo {title} {Social sciences, humanities and the arts for people and the economy}} (\bibinfo {year} {2024}),\ \bibinfo {note} {page Version ID: 1251138241}\BibitemShut {NoStop}%
\bibitem [{\citenamefont {{OECD}}(2015)}]{oecd_frascati_2015}%
  \BibitemOpen
  \bibfield  {author} {\bibinfo {author} {\bibnamefont {{OECD}}},\ }\href {https://www.oecd-ilibrary.org/science-and-technology/frascati-manual-2015_9789264239012-en} {\emph {\bibinfo {title} {Frascati {Manual} 2015: {Guidelines} for {Collecting} and {Reporting} {Data} on {Research} and {Experimental} {Development}}}}\ (\bibinfo  {publisher} {Organisation for Economic Co-operation and Development},\ \bibinfo {address} {Paris},\ \bibinfo {year} {2015})\BibitemShut {NoStop}%
\bibitem [{\citenamefont {of~Statistics}(2020)}]{australian_bureau_of_statistics_australian_2020}%
  \BibitemOpen
  \bibfield  {author} {\bibinfo {author} {\bibfnamefont {A.~B.}\ \bibnamefont {of~Statistics}},\ }\href {https://www.abs.gov.au/statistics/classifications/australian-and-new-zealand-standard-research-classification-anzsrc/latest-release} {\bibinfo {title} {Australian and {New} {Zealand} {Standard} {Research} {Classification} ({ANZSRC}), 2020 {\textbar} {Australian} {Bureau} of {Statistics}}} (\bibinfo {year} {2020})\BibitemShut {NoStop}%
\bibitem [{\citenamefont {Stevenson}\ \emph {et~al.}(2023)\citenamefont {Stevenson}, \citenamefont {Grant}, \citenamefont {Szomszor}, \citenamefont {Ang}, \citenamefont {Kapoor}, \citenamefont {Gunashekar},\ and\ \citenamefont {Guthrie}}]{stevenson_data_2023}%
  \BibitemOpen
  \bibfield  {author} {\bibinfo {author} {\bibfnamefont {C.}~\bibnamefont {Stevenson}}, \bibinfo {author} {\bibfnamefont {J.}~\bibnamefont {Grant}}, \bibinfo {author} {\bibfnamefont {M.}~\bibnamefont {Szomszor}}, \bibinfo {author} {\bibfnamefont {C.}~\bibnamefont {Ang}}, \bibinfo {author} {\bibfnamefont {D.}~\bibnamefont {Kapoor}}, \bibinfo {author} {\bibfnamefont {S.}~\bibnamefont {Gunashekar}},\ and\ \bibinfo {author} {\bibfnamefont {S.}~\bibnamefont {Guthrie}},\ }\href {https://doi.org/10.7249/RRA2162-1} {\emph {\bibinfo {title} {Data enhancement and analysis of the {REF} 2021 {Impact} {Case} {Studies}}}}\ (\bibinfo  {publisher} {RAND Corporation},\ \bibinfo {year} {2023})\BibitemShut {NoStop}%
\bibitem [{\citenamefont {Hook}\ \emph {et~al.}(2018)\citenamefont {Hook}, \citenamefont {Porter},\ and\ \citenamefont {Herzog}}]{hook_dimensions_2018}%
  \BibitemOpen
  \bibfield  {author} {\bibinfo {author} {\bibfnamefont {D.~W.}\ \bibnamefont {Hook}}, \bibinfo {author} {\bibfnamefont {S.~J.}\ \bibnamefont {Porter}},\ and\ \bibinfo {author} {\bibfnamefont {C.}~\bibnamefont {Herzog}},\ }\bibfield  {title} {\bibinfo {title} {Dimensions: {Building} {Context} for {Search} and {Evaluation}},\ }\bibfield  {journal} {\bibinfo  {journal} {Frontiers in Research Metrics and Analytics}\ }\textbf {\bibinfo {volume} {3}},\ \href {https://doi.org/10.3389/frma.2018.00023} {10.3389/frma.2018.00023} (\bibinfo {year} {2018}),\ \bibinfo {note} {publisher: Frontiers}\BibitemShut {NoStop}%
\bibitem [{\citenamefont {Porter}\ \emph {et~al.}(2023)\citenamefont {Porter}, \citenamefont {Hawizy},\ and\ \citenamefont {Hook}}]{porter_recategorising_2023}%
  \BibitemOpen
  \bibfield  {author} {\bibinfo {author} {\bibfnamefont {S.~J.}\ \bibnamefont {Porter}}, \bibinfo {author} {\bibfnamefont {L.}~\bibnamefont {Hawizy}},\ and\ \bibinfo {author} {\bibfnamefont {D.~W.}\ \bibnamefont {Hook}},\ }\bibfield  {title} {\bibinfo {title} {Recategorising research: {Mapping} from {FoR} 2008 to {FoR} 2020 in {Dimensions}},\ }\href {https://doi.org/10.1162/qss_a_00244} {\bibfield  {journal} {\bibinfo  {journal} {Quantitative Science Studies}\ }\textbf {\bibinfo {volume} {4}},\ \bibinfo {pages} {127} (\bibinfo {year} {2023})}\BibitemShut {NoStop}%
\bibitem [{\citenamefont {Jefferson}\ \emph {et~al.}(2018)\citenamefont {Jefferson}, \citenamefont {Jaffe}, \citenamefont {Ashton}, \citenamefont {Warren}, \citenamefont {Koellhofer}, \citenamefont {Dulleck}, \citenamefont {Ballagh}, \citenamefont {Moe}, \citenamefont {DiCuccio}, \citenamefont {Ward}, \citenamefont {Bilder}, \citenamefont {Dolby},\ and\ \citenamefont {Jefferson}}]{jefferson_mapping_2018}%
  \BibitemOpen
  \bibfield  {author} {\bibinfo {author} {\bibfnamefont {O.~A.}\ \bibnamefont {Jefferson}}, \bibinfo {author} {\bibfnamefont {A.}~\bibnamefont {Jaffe}}, \bibinfo {author} {\bibfnamefont {D.}~\bibnamefont {Ashton}}, \bibinfo {author} {\bibfnamefont {B.}~\bibnamefont {Warren}}, \bibinfo {author} {\bibfnamefont {D.}~\bibnamefont {Koellhofer}}, \bibinfo {author} {\bibfnamefont {U.}~\bibnamefont {Dulleck}}, \bibinfo {author} {\bibfnamefont {A.}~\bibnamefont {Ballagh}}, \bibinfo {author} {\bibfnamefont {J.}~\bibnamefont {Moe}}, \bibinfo {author} {\bibfnamefont {M.}~\bibnamefont {DiCuccio}}, \bibinfo {author} {\bibfnamefont {K.}~\bibnamefont {Ward}}, \bibinfo {author} {\bibfnamefont {G.}~\bibnamefont {Bilder}}, \bibinfo {author} {\bibfnamefont {K.}~\bibnamefont {Dolby}},\ and\ \bibinfo {author} {\bibfnamefont {R.~A.}\ \bibnamefont {Jefferson}},\ }\bibfield  {title} {\bibinfo {title} {Mapping the global influence of published research on industry and innovation},\ }\href {https://doi.org/10.1038/nbt.4049} {\bibfield
  {journal} {\bibinfo  {journal} {Nature Biotechnology}\ }\textbf {\bibinfo {volume} {36}},\ \bibinfo {pages} {31} (\bibinfo {year} {2018})},\ \bibinfo {note} {publisher: Nature Publishing Group}\BibitemShut {NoStop}%
\bibitem [{\citenamefont {Jaffe}\ and\ \citenamefont {Trajtenberg}(1996)}]{jaffe_flows_1996}%
  \BibitemOpen
  \bibfield  {author} {\bibinfo {author} {\bibfnamefont {A.}~\bibnamefont {Jaffe}}\ and\ \bibinfo {author} {\bibfnamefont {M.}~\bibnamefont {Trajtenberg}},\ }\bibfield  {title} {\bibinfo {title} {Flows of knowledge from universities and federal laboratories: {Modeling} the flow of patent citations over time and across institutional and geographic boundaries},\ }\href {https://doi.org/10.1073/pnas.93.23.12671} {\bibfield  {journal} {\bibinfo  {journal} {Proceedings of the National Academy of Sciences}\ }\textbf {\bibinfo {volume} {93}},\ \bibinfo {pages} {12671} (\bibinfo {year} {1996})},\ \bibinfo {note} {publisher: Proceedings of the National Academy of Sciences}\BibitemShut {NoStop}%
\bibitem [{\citenamefont {Nye}(2004)}]{nye_soft_2004}%
  \BibitemOpen
  \bibfield  {author} {\bibinfo {author} {\bibfnamefont {J.~S.}\ \bibnamefont {Nye}},\ }\href@noop {} {\emph {\bibinfo {title} {Soft power: the means to success in world politics}}},\ \bibinfo {edition} {1st}\ ed.\ (\bibinfo  {publisher} {Public Affairs},\ \bibinfo {address} {New York},\ \bibinfo {year} {2004})\ \bibinfo {note} {oCLC: 53951715}\BibitemShut {NoStop}%
\bibitem [{\citenamefont {Nye}(2008)}]{nye_public_2008}%
  \BibitemOpen
  \bibfield  {author} {\bibinfo {author} {\bibfnamefont {J.~S.}\ \bibnamefont {Nye}},\ }\bibfield  {title} {\bibinfo {title} {Public {Diplomacy} and {Soft} {Power}},\ }\href {https://doi.org/10.1177/0002716207311699} {\bibfield  {journal} {\bibinfo  {journal} {The ANNALS of the American Academy of Political and Social Science}\ }\textbf {\bibinfo {volume} {616}},\ \bibinfo {pages} {94} (\bibinfo {year} {2008})},\ \bibinfo {note} {publisher: SAGE Publications Inc}\BibitemShut {NoStop}%
\bibitem [{\citenamefont {Hook}\ and\ \citenamefont {Wilsdon}(2023)}]{hook_pandemic_2023}%
  \BibitemOpen
  \bibfield  {author} {\bibinfo {author} {\bibfnamefont {D.~W.}\ \bibnamefont {Hook}}\ and\ \bibinfo {author} {\bibfnamefont {J.~R.}\ \bibnamefont {Wilsdon}},\ }\bibfield  {title} {\bibinfo {title} {The pandemic veneer: {COVID}-19 research as a mobilisation of collective intelligence by the global research community},\ }\href {https://doi.org/10.1177/26339137221146482} {\bibfield  {journal} {\bibinfo  {journal} {Collective Intelligence}\ }\textbf {\bibinfo {volume} {2}},\ \bibinfo {pages} {26339137221146482} (\bibinfo {year} {2023})},\ \bibinfo {note} {publisher: SAGE Publications}\BibitemShut {NoStop}%
\bibitem [{\citenamefont {Hook}\ and\ \citenamefont {Porter}(2025)}]{porterhook2025}%
  \BibitemOpen
  \bibfield  {author} {\bibinfo {author} {\bibfnamefont {D.~W.}\ \bibnamefont {Hook}}\ and\ \bibinfo {author} {\bibfnamefont {S.~J.}\ \bibnamefont {Porter}},\ }\bibfield  {title} {\bibinfo {title} {In preparation},\ }\href@noop {} {\  (\bibinfo {year} {2025})}\BibitemShut {NoStop}%
\end{thebibliography}%

\end{document}